\documentclass[twocolumn]{aastex6}
\usepackage{graphicx}

\usepackage{placeins}
\usepackage{amsmath}

\newcommand{\kms}{km s$^{-1}$}

\begin{document}

\title{Disk-Like Chemistry of the Triangulum-Andromeda Overdensity as Seen by APOGEE}

\author{Christian R. Hayes\altaffilmark{1}, Steven R. Majewski\altaffilmark{1}, Sten Hasselquist\altaffilmark{2}, Rachael L. Beaton\altaffilmark{3,4,5,6}, Katia Cunha\altaffilmark{7,8}, Verne V. Smith\altaffilmark{9}, Adrian M. Price-Whelan\altaffilmark{5}, Borja Anguiano\altaffilmark{1,10}, Timothy C. Beers\altaffilmark{11}, Ricardo Carrera\altaffilmark{12}, J. G. Fern\'{a}ndez-Trincado\altaffilmark{13,14}, Peter M. Frinchaboy\altaffilmark{15}, D. A. Garc\'{i}a-Hern\'{a}ndez\altaffilmark{16,17}, Richard R. Lane\altaffilmark{18,29}, David L. Nidever\altaffilmark{9,20}, Christian Nitschelm\altaffilmark{21}, Alexandre Roman-Lopes\altaffilmark{22}, Olga Zamora\altaffilmark{16,17}}
\altaffiltext{1}{Department of Astronomy, University of Virginia, Charlottesville, VA 22904-4325, USA}
\altaffiltext{2}{Department of Astronomy, New Mexico State University, Las Cruces, NM 88003, USA}
\altaffiltext{3}{Hubble Fellow}
\altaffiltext{4}{Carnegie-Princeton Fellow}
\altaffiltext{5}{Department of Astrophysical Sciences, Princeton University, 4 Ivy Lane, Princeton, NJ~08544}
\altaffiltext{6}{The Observatories of the Carnegie Institution for Science, 813 Santa Barbara St., Pasadena, CA~91101}
\altaffiltext{7}{Observat\'{o}rio Nacional, 77 Rua General Jos\'{e} Cristino, Rio de Janeiro, 20921-400, Brazil}
\altaffiltext{8}{Steward Observatory, University of Arizona, 933 North Cherry Avenue, Tucson, AZ 85721, USA}
\altaffiltext{9}{National Optical Astronomy Observatory, 950 North Cherry Avenue, Tucson, AZ 85719, USA}
\altaffiltext{10}{Department of Physics and Astronomy, Macquarie University, North Ryde, NSW 2109, Australia}
\altaffiltext{11}{Department of Physics, University of Notre Dame, and JINA Center for the Evolution of the Elements, Notre Dame, IN 46556 USA}
\altaffiltext{12}{Astronomical Observatory of Padova, National Institute of Astrophysics, Vicolo Osservatorio 5 - 35122 - Padova}
\altaffiltext{13}{Departamento de Astronom\'{i}a, Universidad de Concepci\'{o}n, Casilla 160-C, Concepci\'{o}n, Chile}
\altaffiltext{14}{Institut Utinam, CNRS UMR6213, Univ. Bourgogne Franche-Comt\'e, OSU THETA , Observatoire de Besan\c{c}on, BP 1615, 25010 Besan\c{c}on Cedex, France}
\altaffiltext{15}{Department of Physics and Astronomy, Texas Christian University, Fort Worth, TX 76129}
\altaffiltext{16}{Instituto de Astrof\'{i}sica de Canarias (IAC), V\'{i}a L\'{a}ctea, E-38205 La Laguna, Tenerife, Spain}
\altaffiltext{17}{Departamento de Astrof\'{i}sica, Universidad de La Laguna (ULL), E-38206 La Laguna, Tenerife, Spain}
\altaffiltext{18}{Millennium Institute of Astrophysics, Av. Vicu\~{n}a Mackenna 4860, 782-0436 Macul, Santiago, Chile}
\altaffiltext{19}{Instituto de Astrof\'{i}sica, Pontificia Universidad Cat\'{o}lica de Chile, Av. Vicu\~{n}a Mackenna 4860, 782-0436 Macul, Santiago, Chile}
\altaffiltext{20}{Department of Physics, Montana State University, P.O. Box 173840, Bozeman, MT 59717-3840}
\altaffiltext{21}{Unidad de Astronom\'{i}a, Universidad de Antofagasta, Avenida Angamos 601, Antofagasta 1270300, Chile}
\altaffiltext{22}{Departamento de F\'{i}sica, Facultad de Ciencias, Universidad de La Serena, Cisternas 1200, La Serena, Chile}
\email{crh7gs@virginia.edu}

\begin{abstract}

The nature of the Triangulum-Andromeda (TriAnd) system has been debated since the discovery of this distant, low-latitude Milky Way (MW) overdensity more than a decade ago.  Explanations for its origin are either as a halo substructure from the disruption of a dwarf galaxy or a distant extension of the Galactic disk.  We test these hypotheses using chemical abundances of a dozen TriAnd members from the Sloan Digital Sky Survey's 14th Data Release of Apache Point Observatory Galactic Evolution Experiment (APOGEE) data to compare to APOGEE abundances of stars with similar metallicity from both the Sagittarius (Sgr) dSph, and the outer MW disk.  We find that TriAnd stars are chemically distinct from Sgr across a variety of elements, (C+N), Mg, K, Ca, Mn, and Ni, with a separation in [X/Fe] of about 0.1 to 0.4 dex depending on the element.  Instead, the TriAnd stars, with a median metallicity of about $-$0.8, exhibit chemical abundance ratios similar to those of the lowest metallicity ([Fe/H] $\sim -0.7$) stars in the outer Galactic disk, and are consistent with expectations of extrapolated chemical gradients in the outer disk of the MW.  These results suggest that TriAnd is associated with the MW disk, and, therefore, that the disk extends to this overdensity --- i.e., past a Galactocentric radius of 24 kpc --- albeit vertically perturbed about 7 kpc below the nominal disk midplane in this region of the Galaxy.

\end{abstract}

\keywords{ Galaxy:  structure $-$ Galaxy:  evolution $-$ Galaxy:  disk $-$ Galaxy:  halo $-$ stars:  abundances }

\section{Introduction}
	
Several overdensities discovered towards the outer disk of the Milky Way (MW) have origins that are still not understood, including ``Triangulum-Andromeda'' \citep[TriAnd,][]{maj04,rp04}, which is a low-latitude, distant ($\sim 20$ kpc), and kinematically cold ($\sigma_{\rm LOS} \sim 25$ km s$^{-1}$) cloud of stars \citep{she14}.  Theories to explain TriAnd's observed properties include that it (1) could be tidal debris from a disrupted dwarf galaxy \citep{dea14,she14}, or (2) represents part of an extended and perturbed MW disk, perhaps a trough in a series of midplane oscillations \citep{pw15,xu15,li17}.  Recent simulations have illustrated that large, non-axisymmetric, vertical oscillations can be excited in the outer disk due to the interaction of the Sagittarius dwarf spheroidal galaxy (Sgr dSph) with the MW; reproducing structures reminiscent of TriAnd \citep[and other overdensities;][]{lap17}. These differing origin scenarios should impart different chemical signatures to TriAnd stars.  For example, if TriAnd is the result of a perturbation to the outer Galactic disk, its chemical abundance patterns should resemble that of known outer disk stars, whereas tidal debris should share the chemistry seen in dwarf galaxies.

To date, chemical studies of TriAnd have reached differing conclusions about its origin. The first high-resolution spectroscopic study of TriAnd stars by \citet{chou11} focused on the elements Ti, Y, and La, and found TriAnd had some chemical differences from MW disk stars in the solar neighborhood, and suggested a dwarf galaxy origin. On the other hand, recent study of O, Na, Mg, Ti, Ba, and Eu abundance ratios in TriAnd stars indicate that it is chemically consistent with the MW disk rather than a dwarf galaxy \citep{ber18}.  Further chemical study of TriAnd to resolve such discrepancies is clearly warranted.

The Apache Point Observatory Galactic Evolution Experiment \citep[APOGEE,][]{apogee} provides such an opportunity.  APOGEE is a high-resolution ($R \sim 22,500$) spectroscopic survey of Galactic stellar populations with $H$-band sensitivity well-suited to the exploration of highly extinguished low-latitude targets, such as the TriAnd overdensity and the outer disk.  Selecting from the $\sim$263,000 stars observed with the SDSS 2.5-m telescope \citep{gunn06} and analyzed by APOGEE in the 14th Data Release \citep[DR14,][]{dr14} of the Sloan Digital Sky Survey-IV \citep[SDSS-IV,][]{sdss4}, we use the abundances of six APOGEE-measured elements to compare TriAnd red giants to outer disk and Sgr dSph stars, and demonstrate that the TriAnd chemistry is more consistent with an extrapolation of outer MW disk chemical gradients than the abundance patterns of a prototypical dwarf galaxy of similar enrichment.
	
\section{Data}

Details of the APOGEE survey and data reduction pipeline can be found in \citet{apogee} and \citet{dln15}, respectively.  Here we use the SDSS-IV DR14 calibrated stellar parameters and chemical abundances derived from the APOGEE Stellar Parameter and Chemical Abundance Pipeline \citep[ASPCAP;][]{aspcap}.  To insure that we are considering the most reliably determined stellar parameters, we remove stars flagged\footnote{A description of these flags can be found in the online SDSS DR14 bitmask documentation (\url{http://www.sdss.org/dr14/algorithms/bitmasks/})} with the {\sc starflags} \textsc{bad\_pixels}, \textsc{very\_bright\_neighbor}, or \textsc{low\_snr} set, or any stars with the \textsc{aspcapflags},   \textsc{rotation\_warn} or \textsc{star\_bad}.  We also restrict analysis to stars with small visit-to-visit velocity scatter, $V_{\rm scatter} \leq 1$ \kms, low velocity uncertainty, $V_{\rm err} \leq 0.2$ \kms, and S/N $> 80$, to remove stars whose ASPCAP-analyzed spectra may be of lower quality.  Finally, we focus on stars in effective temperature ranges, between 3700 K and 5500 K, where stellar parameters and chemical abundances are reliably and consistently determined.

In this high-quality sample, we have 12 M giants that were identified by \citet{she14} as TriAnd members from their photometry and cold kinematics ($\sigma \sim 25$ km s$^{-1}$) and were deliberately targeted in APOGEE-2 \citep{zas17}.  The APOGEE-measured properties of these TriAnd stars are given in Table \ref{table1}.  Several studies have suggested that TriAnd may separate into two features, TriAnd1 and TriAnd2, that coincide on the sky but lie at photometrically determined heliocentric distances $\sim 20$ kpc and $\sim 28$ kpc, respectively \citep{mar07}.  However, since these features were shown to overlap considerably in spectrophotometric distance, radial velocity, and metallicity \citep{she14}, and we only have two stars classified as TriAnd2 members, we treat them here as a single overdensity. For a comparison, we use Sgr dSph because this dwarf galaxy is sufficiently enriched to have a considerable M giant population, like TriAnd.  To do so, we use a set of 69 Sgr dSph members confirmed by \citet{has17} and satisfy our quality criteria.

\begin{deluxetable*}{c l p{4.5cm} c l p{4.5cm}}
\tabletypesize{\scriptsize}
\tablewidth{0pt}
\tablecolumns{3}
\tablecaption{Properties of TriAnd Stars \label{table1}}
\tablehead{\colhead{Column} & \colhead{Column Label} & \colhead{Column Description} & \colhead{Column} & \colhead{Column Label} & \colhead{Column Description}}
\startdata
1 & APOGEE & APOGEE Star ID & 17 & Vturb & Microturbulent velocity (km s$^{-1}$) \\
2 & RAdeg & Right Ascension (decimal degrees) & 18 & Vmacro & Macroturbulent velocity (km s$^{-1}$) \\
3 & DEdeg & Declination (decimal degrees) & 19 & [Fe/H] & Log abundance, [Fe/H] \\
4 & GLON & Galactic Longitude (decimal degrees) & 20 & e\_[Fe/H] & Uncertainty in [Fe/H] \\
5 & GLAT & Galactic Latitude (decimal degrees) & 21 & [CN/Fe] & Log abundance, [(C+N)/Fe] \\
6 & Jmag & 2MASS J magnitude & 22 & e\_[CN/Fe] & Uncertainty in [(C+N)/Fe] \\
7 & Hmag & 2MASS H magnitude & 23 & [Mg/Fe] & Log abundance, [Mg/Fe] \\
8 & Kmag & 2MASS Ks magnitude & 24 & e\_[Mg/Fe] & Uncertainty on [Mg/Fe] \\
9 & Dist & Heliocentric distance (kpc) & 25 & [K/Fe] & Log abundance, [K/Fe] \\
10 & e\_Dist & Uncertainty in distance (kpc) & 26 & e\_[K/Fe] & Uncertainty on [K/Fe] \\
11 & HRV & Heliocentric radial velocity (km s$^{-1}$) & 27 & [Ca/Fe] & Log abundance, [Ca/Fe] \\
12 & e\_HRV & Radial velocity uncertainty (km s$^{-1}$) & 28 & e\_[Ca/Fe] & Uncertainty on [Ca/Fe] \\
13 & Teff& Effective surface temperature (K) & 29 & [Mn/Fe] & Log abundance, [Mn/Fe] \\
14 & e\_Teff & Uncertainty in $T_{\rm eff}$ (K) & 30 & e\_[Mn/Fe] & Uncertainty on [Mn/Fe] \\
15 & logg & Surface gravity & 31 & [Ni/Fe] & Log abundance, [Ni/Fe]\\
16 & e\_logg & Uncertainty in $\log g$ & 32 & e\_[Ni/Fe] & Uncertainty on [Ni/Fe] \\
\enddata
\tablecomments{Table 1 is published in its entirety in the machine-readable format.  A portion is shown here for guidance regarding its form and content.}
\tablecomments{Null entries are given values of -9999.}
\end{deluxetable*}

\begin{figure*}
  \centering
  \includegraphics[scale=0.4,trim = 1.25in 0in 1.25in 0in, clip]{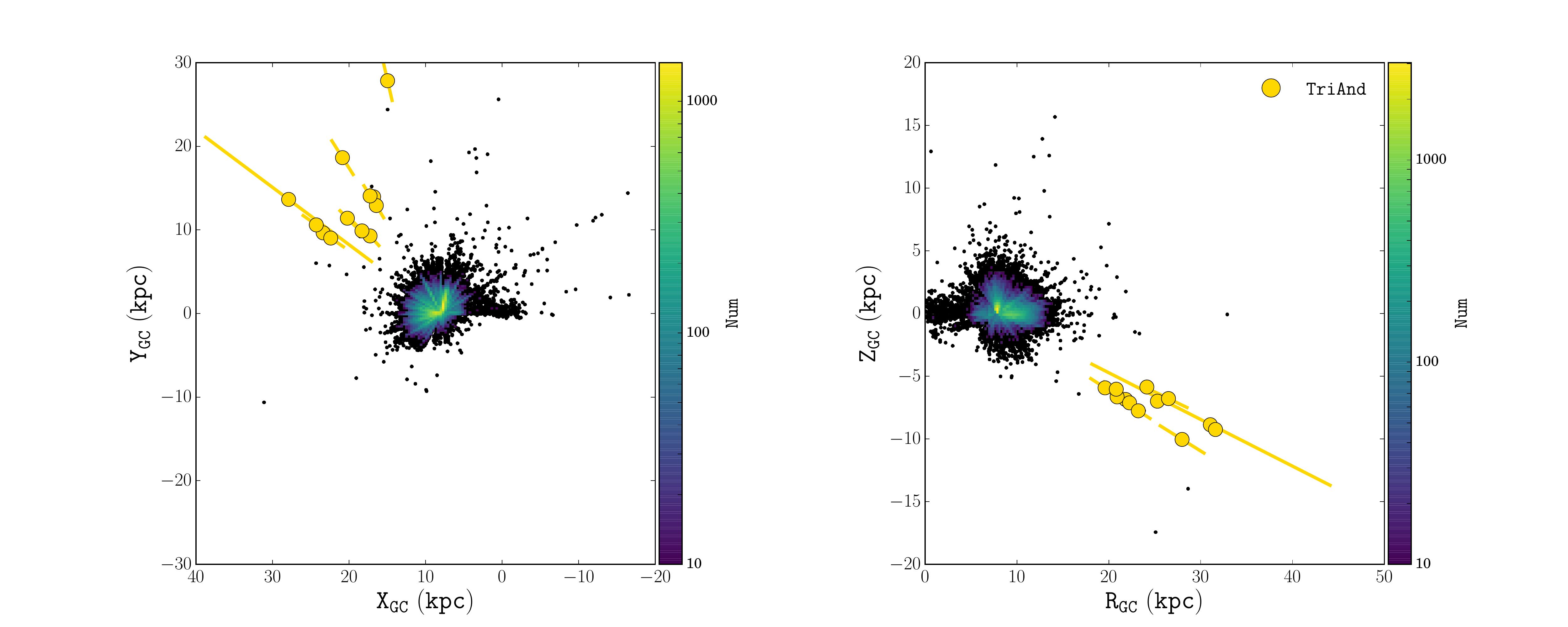}
  \caption{Spatial distribution of the high-quality APOGEE stars (before outer disk selection, black points) and TriAnd stars (gold circles), showing the reported $1\sigma$ distance uncertainty for the TriAnd stars. (left) Stellar distribution projected onto the Galactic plane. (right) Distribution azimuthally collapsed onto the cylindrical $R_{\rm GC}$-$Z_{\rm GC}$ plane.}
  \label{dist}
\end{figure*}

We also compile a representative comparison sample of outer disk stars from APOGEE.  To do so, we use spectrophotometric distances calculated by \citet{starhorse} using DR14 ASPCAP stellar parameters and their Bayesian \texttt{StarHorse} code.  Because not all of these distances are reliable, we only use stars that are not flagged with {\sc high\_extinction\_warn}, {\sc nummodels\_bad}, or {\sc extinction\_bad\_bright2mass}, which may have erroneous distance estimates due to poor extinction corrections or lack available stellar models to determine a reliable distance.  Because we will determine metallicity and chemical gradients in the outer disk to compare to the chemistry of TriAnd stars, we also want stars with relatively accurate \texttt{StarHorse} distances, and remove stars with $\sigma_d > 0.5$ kpc on their posterior distance distribution.  These distances are converted to Galactocentric coordinates assuming $R_{{\rm GC}, \sun} = 8$ kpc, and the Galactic distribution of these APOGEE stars is shown in Figure \ref{dist}, along with the TriAnd stars with \texttt{StarHorse} distances.  Finally, we form our ``outer disk sample'' from this high-quality set of MW stars by selecting those with cylindrical Galactocentric radii $R_{\rm GC} > 9$ kpc and midplane distances $|Z| < 1.0$ kpc.

\section{Results and Analysis}

Using the reliable spectrophotometric distances from \texttt{StarHorse}, our TriAnd sample (with median distance uncertainties of 2 kpc) is centered at a median distance of $\sim$18 kpc with a $1\sigma$ spread of 4 kpc.  This is consistent with past distances found for TriAnd, e.g., the 18.2 kpc distance \citep{she14} used to select the ``TriAnd1'' members that dominate our sample here.  This puts the TriAnd sample at a median Galactocentric radius of $\sim$24 kpc (1-$\sigma_{\rm R}$ spread of 4 kpc) and below the disk by $\sim$7 kpc (1-$\sigma_{\rm Z}$ spread of 1 kpc).  

\begin{figure}
  \centering
  \includegraphics[scale=0.4]{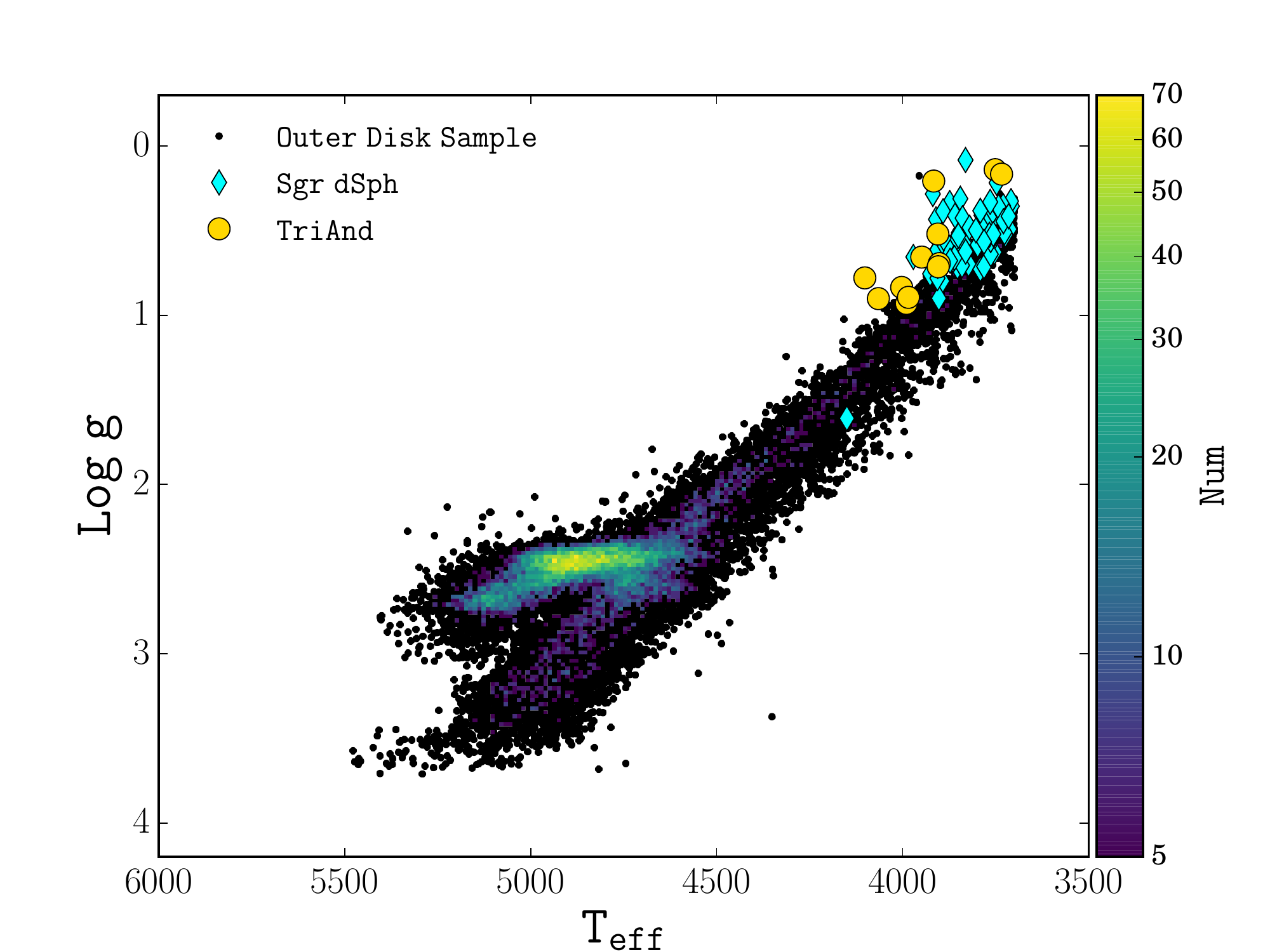}
  \caption{Spectroscopic HR diagram of APOGEE-derived $\log g$ versus $T_{\rm eff}$ for the outer disk (black points with a 2D histogram where densely populated), TriAnd (gold circles), and Sgr dSph (cyan diamonds) samples.}
  \label{hrd}
\end{figure}

While a few of the TriAnd stars do not have reliable spectrophotometric distances from \texttt{StarHorse}, they were selected by \citet{she14} in color-magnitude to fall along the red giant branches of 8 Gyr/10 Gyr $-0.8$/$-1.0$ metallicity isochrones at heliocentric distances around 18.2 kpc/27.5 kpc for TriAnd1/TriAnd2. Figure \ref{hrd} shows that TriAnd stars have effective temperatures and surface gravities of cool red giants, supporting the isochrone-based distances used by \citet{she14}.

As shown below, we can perform a more robust analysis by comparing to a large sample of outer disk stars spanning a considerable range of Galactocentric radii.  To achieve this, we have not restricted the outer disk sample to cover the $T_{\rm eff}$ and $\log g$ range of the TriAnd and Sgr dSph samples.  However, to ensure that this does not affect our chemical abundance comparison, we examined the abundance distributions of relevant chemical elements for outer disk stars warmer and cooler than 4250 K to verify that there were no significant differences in their abundance patterns at a level that affects our conclusions about TriAnd.  Because our TriAnd and Sgr dSph samples cover nearly the same stellar parameter space, their comparison should be even more robustly reliable.

\begin{figure*}
  \centering
  \includegraphics[scale=0.4,trim = 1.5in 0.3in 1.5in 1.0in, clip]{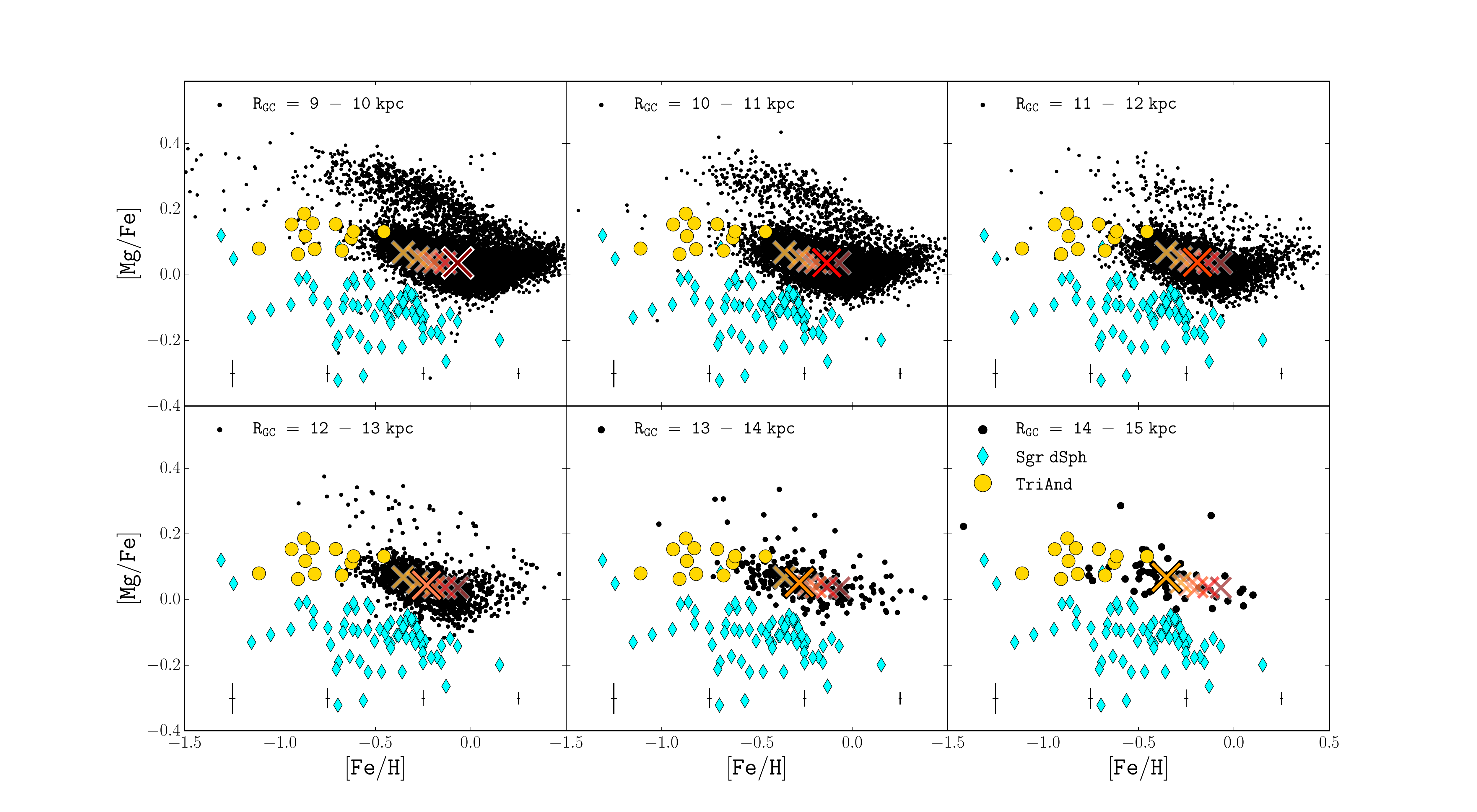}
  \caption{The [Mg/Fe]-[Fe/H] distributions in 1 kpc wide Galactocentric annular bins for the outer disk (black points, with increasing Galactocentric radius from the top left to the bottom right), TriAnd (gold circles) and Sgr dSph (cyan diamonds) samples.  The median in each outer disk annulus has been marked with a cross, colored chromatically from dark red to orange for the innermost to the outermost annuli, with the medians in other annuli shown as smaller crosses of their respective colors.  The plotted error bars show the median internal abundance uncertainties in 0.5 dex wide metallicity bins.} 
  \label{mgdist}
\end{figure*}

Despite its large size (21,868 stars), our APOGEE-based disk sample \citep[targeted to minimize selection biases;][]{zas13,zas17} does not extend to the distance of the TriAnd stars and has few stars beyond $R_{\rm GC} > 15$ kpc.  However, if the abundances of Galactic disk stars follow relatively well-behaved radial metallicity and chemical gradients, we can extrapolate those trends to establish the abundances expected for the disk at the distance of TriAnd.  To illustrate this, Figure \ref{mgdist} shows the [Mg/Fe]-[Fe/H] plane for outer disk stars with [Mg/Fe] and [Fe/H] uncertainties less than 0.1 dex, subdivided into samples lying within 1 kpc wide annuli spanning Galactocentric radii from $R_{\rm GC} = 9$ kpc to 15 kpc.  Within each annulus we calculate the median abundance of the outer disk sample, and can see that there are clear trends in both [Fe/H] and [Mg/Fe] with Galactocentric radius.

Figure \ref{mgdist} strikingly demonstrates that the [Mg/Fe]-[Fe/H] distribution for TriAnd stars occupies a region of this parameter space consistent with a metal-poor extrapolation of the outer disk trend to larger radius.  Moreover, the TriAnd stars are enhanced in [Mg/Fe] relative to Sgr dSph stars of similar metallicity.

\begin{figure*}
  \centering
  \includegraphics[scale=0.4,trim = 1.25in 1.6in 1.25in 2.in, clip]{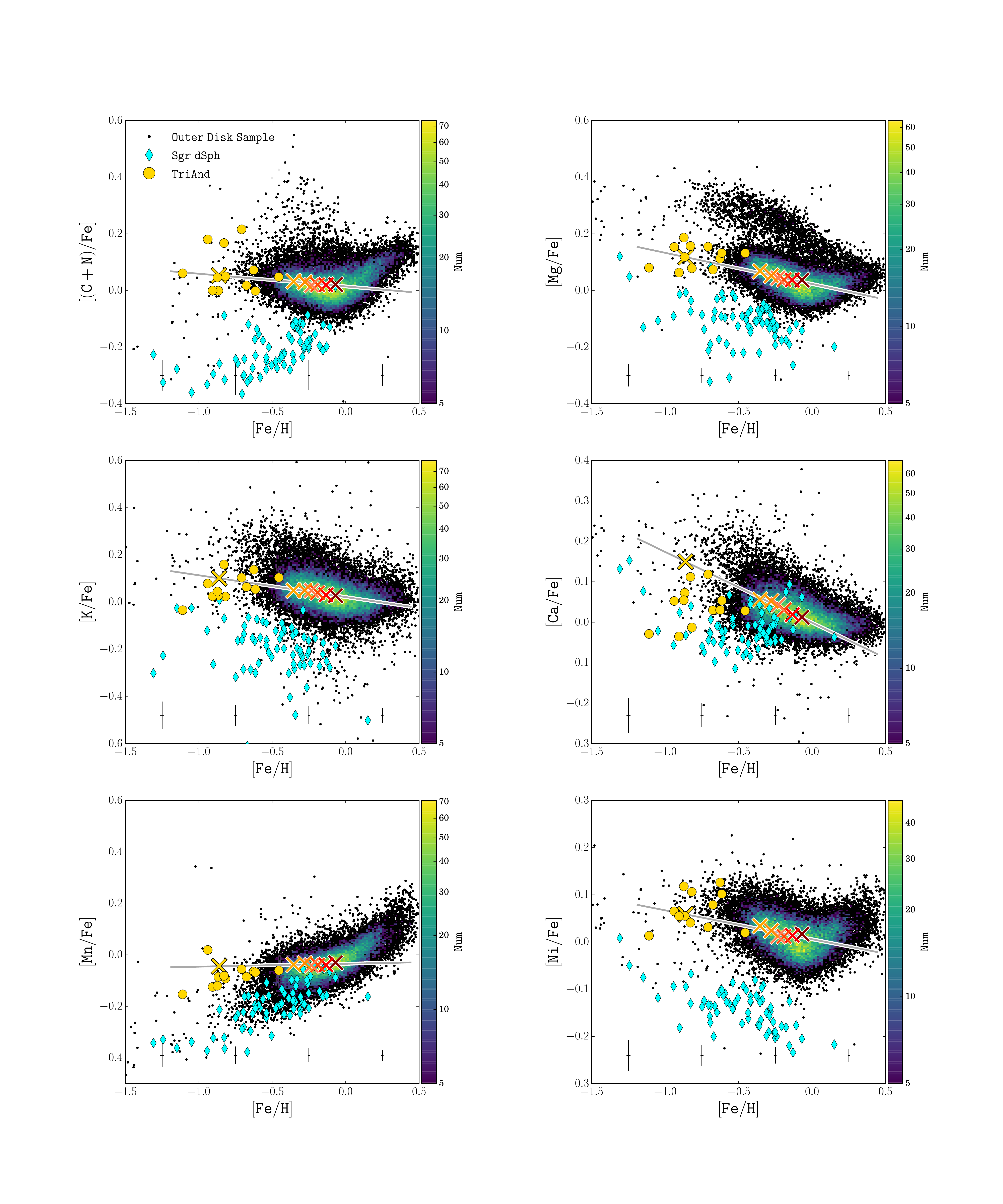}
  \caption{[X/Fe] (for (C+N), Mg, K, Ca, Mn, and Ni) versus [Fe/H] for the outer disk, TriAnd, and Sgr dSph samples (with colors and symbols as in Figure \ref{hrd}).  The medians of the outer disk sample in 1 kpc wide Galactocentric annuli are plotted as crosses, colored according to Galactocentric radius, as in Figure \ref{mgdist}.  Linear parametric fits to [X/Fe]-[Fe/H] medians from $R_{\rm GC}$ of 9 kpc to 15 kpc as a function of $R_{\rm GC}$ are shown (gray lines) and extrapolated out to 30 kpc, with $R_{\rm GC} = $ 24 kpc (the median Galactocentric radius of our TriAnd sample) marked as a gold cross. Typical uncertainties are shown as in Figure \ref{mgdist}.}
  \label{multi_elem}
\end{figure*}

APOGEE enables similar comparisons of these samples in multiple  chemical dimensions.  Figure \ref{multi_elem} presents the chemical abundance distributions of the TriAnd, outer disk, and Sgr dSph samples (showing only stars with both $\sigma$[X/Fe] and $\sigma$[Fe/H] $< 0.1$ dex) for a set of elements are formed in a variety of nucleosynthetic processes: the $\alpha$-elements Mg and Ca, the odd-$Z$ element K, the iron-peak elements Mn and Ni, and the sum of C and N \citep[surface abundances of C and N are altered during dredge-up and mixing in red giants, but their sum is effectively conserved; see][and references therein]{mar16}. This subset of APOGEE-measured elements were specifically chosen because they do not exhibit different abundance trends in warm and cool outer disk stars, and are measured with low uncertainties.

In these other chemical planes, as in Figure \ref{mgdist}, the TriAnd stars tend to overlap and extend the sequence of the outer disk stars at metallicities of [Fe/H] $\sim$$-0.7$.  Moreover, the radial gradient of the outer disk, as measured by the median chemistry in 1 kpc annular rings (also as in Figure \ref{mgdist}), approaches the chemical abundances of TriAnd stars at increasing radii. We can extrapolate these trends to the distance of TriAnd to estimate the expected abundances for outer disk stars at that location.  While the shape of the [X/Fe]-[Fe/H] distribution of the outer disk sample is complex within each annulus, as seen in Figure \ref{mgdist}, the median chemistry of the distributions varies roughly linearly with Galactocentric radius.  Thus, we fit the annular median abundances linearly in Galactocentric radius to find a metallicity gradient of $\partial$[Fe/H]$/\partial R_{\rm GC} = -0.051 \pm 0.005$ dex kpc$^{-1}$ and $\partial$[X/Fe]$/\partial R_{\rm GC}$ gradients for ((C+N), Mg, K, Ca, Mn, Ni) $= (0.002 \pm 0.001, \ 0.006 \pm 0.001, \ 0.005 \pm 0.001, \ 0.009 \pm 0.001, \ -0.001 \pm 0.001, \ 0.003 \pm 0.002) $ dex kpc$^{-1}$.

\section{Discussion}

Figure \ref{multi_elem} demonstrates that, when extrapolated to the distance of TriAnd, the outer disk chemistry generally matches that of the TriAnd stars, which suggests that they are associated.  For [Ca/Fe] the predicted median disk chemistry lies at the edge of the TriAnd distribution, which appears to indicate that a {\it linear} extrapolation of the disk Ca gradient may not be appropriate.  We note that the flattening of the [Ca/Fe] trend in the outer disk at the largest radii seems astrophysically significant, and in comparison with the nearly constant gradient in the lower mass alpha-element, Mg, may be reflecting radial or time variations in the initial mass function or star formation history of the disk.

In contrast to the agreement with radial extrapolations of outer disk chemistry, the abundance patterns of the TriAnd stars are distinct from those of Sgr dSph stars, despite their similar metallicities.  If the Sgr dSph is representative of the chemistry of relatively enriched tidal debris falling into the MW, then the chemical differences between TriAnd and Sgr dSph supports the notion that TriAnd is not tidal debris, at least from this type of dwarf galaxy.  Another example of a relatively enriched dwarf galaxy is the Large Magellanic Cloud (LMC).  While the $\alpha$-element abundances of the LMC overlap some with the metal-poor stars in the thin disk (and thus with TriAnd), the LMC exhibits low Ni abundances, with [Ni/Fe] $\sim -0.2$ \citep{vds13}, about 0.3 dex lower than the [Ni/Fe] ratios found here in TriAnd stars.  Thus, the LMC provides another example of a dwarf galaxy with distinct chemistry from TriAnd.

What about other potentially major accretion sources? Studies of metal-poor stars have uncovered two chemically distinct populations in the MW \citep[e.g.,][]{ns10,haw15,hay18}. These two populations are \citep[using the definitions from][]{hay18,fernandez18} a ``low-magnesium'' halo population, thought to be accreted satellite galaxy debris, and a ``high-magnesium'' population, which continues the chemical trends of the thick disk and has the chemistry expected of the classical halo.  The [X/Fe] ratios of TriAnd stars in (C+N), K, Mn, and Ni are $0.2$-$0.4$ dex higher than the population of MW field stars thought to be accreted halo stars, and are also differentiated from the more classical halo population (and thick disk), which have high $\alpha$-element abundances.

The results here are in agreement with those of \citet{ber18}, who argue for an association of TriAnd with the MW disk based on O, Na, Mg, Ti, Ba, and Eu abundances of eight TriAnd stars\footnote{We have two stars in common with \citeauthor{ber18}, and the derived properties agree between the two studies within uncertainties and offsets of a typical size observed between APOGEE and optical studies \citep{jon18}.} compared to MW, Fornax dSph, and Sgr dSph star samples.  However, our conclusions are at odds with those reached by \citet{chou11}, who proposed that TriAnd is more likely to be debris from a disrupted dwarf galaxy.  

\citet{chou11} found a {\it mean} [Ti/Fe] ratio in their TriAnd sample of six stars (none of which overlap the \citeauthor{ber18} sample or ours) about 0.2 dex lower than in Sgr dSph stars, and significantly lower than for their sample of MW stars from the solar neighborhood.  These findings led to the original conclusion that TriAnd enriched slower than either population, consistent with expectations for a slowly enriching dwarf galaxy.  However, this conclusion was largely drawn because half (three) of the stars in the \citeauthor{chou11} TriAnd sample had [Ti/Fe] $\sim 0.5$ dex lower than their MW comparison sample despite (a) the remainder of their sample having [Ti/Fe] ratios consistent with the MW disk sample, and (b) \citeauthor{chou11} finding their TriAnd sample to have $s$-process abundances in La and Y consistent with their MW disk trend.  Unfortunately Ti abundances cannot be reliably measured by ASPCAP currently \citep{haw16,sou16}, and we cannot study TriAnd Ti abundances here.

Given the disk-like $\alpha$-element abundances found for TriAnd stars by \citet[][including disk-like Ti abundances]{ber18}, and those found for our TriAnd sample here, it seems that the lower [Ti/Fe] ratios found by \citet{chou11} may not be representative of the $\alpha$-element abundances of TriAnd as a whole.  Instead, the apparently considerably lower [Ti/Fe] in three \citet{chou11} stars relative to the disk may be due to a variety of causes, including random measurement errors, systematic offsets between their TriAnd [Ti/Fe] and their adopted disk chemistries from the literature, or small number statistics drawing from a TriAnd population with a potentially large intrinsic scatter in Ti abundances.  Reconsidering that the Y, La and half of the Ti abundances in the \citeauthor{chou11} sample are consistent with MW disk abundance patterns, their results could be reinterpreted as supporting a disk origin.

In summary, we find that TriAnd is chemically distinct from the Sgr dSph, having [X/Fe] $\sim 0.1-0.4$ dex higher in (C+N), Mg, K, Ca, Mn, and Ni, and is also distinct from the LMC in its Ni abundances, having [Ni/Fe] $\sim$ 0.3 dex higher than the LMC stars observed by \citet{vds13}.  On the other hand, while our TriAnd stars are typically more metal-poor ([Fe/H] $\sim -0.8$) than most outer disk stars sampled by APOGEE, TriAnd does appear to overlap in chemical space with the lowest metallicity stars ([Fe/H] $\sim -0.7$) known to lie in the outer regions of the disk.  Moreover, linear extrapolation of each of the chemical gradients measured in the APOGEE outer disk sample to the Galactocentric radii of the TriAnd stars predicts abundances similar to those found in our TriAnd sample.  These results support the proposition that TriAnd is associated with the outer disk of the MW and, if so, its large distance from the midplane ($\sim 7$ kpc) may be the result of a perturbation to the MW disk \citep[as in][]{lap17}.

If the TriAnd overdensity is indeed a feature of the MW disk, then that would imply that the disk extends to radii $\gtrsim$ $24$ kpc (i.e., the Galactocentric radius of our TriAnd sample), as suggested by \citet{lop18}.  By inference, this greater MW disk would extend through the radii occupied by other Galactic anticenter overdensities (such as the Monoceros Ring) and lend greater weight to the notion that they, too, are parts of the MW disk.

\acknowledgements
The authors thank the referee for his/her helpful comments.  This research used {\sc topcat} \citep{topcat}.  CRH acknowledges the NSF Graduate Research Fellowship through Grant DGE-1315231.  CRH and SRM acknowledge NSF grants AST-1312863 and AST-1616636. Support for this work was provided by NASA through Hubble Fellowship grant \#51386.01 awarded to RLB by the Space Telescope Science Institute, which is operated by the Association of  Universities for Research in Astronomy, Inc., for NASA, under contract NAS 5-26555.  DAGH and OZ acknowledge support provided by the Spanish Ministry of Economy and Competitiveness (MINECO) under grant AYA-2017-88254-P.  TCB acknowledges partial support from grant PHY 14-30152 (Physics Frontier Center/JINA-CEE), awarded by the U.S. National Science Foundation.  J.G.F-T is supported by FONDECYT No. 3180210.

Funding for the Sloan Digital Sky Survey IV has been provided by the Alfred P. Sloan Foundation, the U.S. Department of Energy Office of Science, and the Participating Institutions. SDSS-IV acknowledges
support and resources from the Center for High-Performance Computing at the University of Utah. The SDSS web site is www.sdss.org.

SDSS-IV is managed by the Astrophysical Research Consortium for the 
Participating Institutions of the SDSS Collaboration including the 
Brazilian Participation Group, the Carnegie Institution for Science, 
Carnegie Mellon University, the Chilean Participation Group, the French Participation Group, Harvard-Smithsonian Center for Astrophysics, Instituto de Astrof\'isica de Canarias, The Johns Hopkins University, Kavli Institute for the Physics and Mathematics of the Universe (IPMU)/University of Tokyo, Lawrence Berkeley National Laboratory, Leibniz Institut f\"ur Astrophysik Potsdam (AIP), Max-Planck-Institut f\"ur Astronomie (MPIA Heidelberg), Max-Planck-Institut f\"ur Astrophysik (MPA Garching), Max-Planck-Institut f\"ur Extraterrestrische Physik (MPE), National Astronomical Observatories of China, New Mexico State University, New York University, University of Notre Dame, Observat\'ario Nacional/MCTI, The Ohio State University, Pennsylvania State University, Shanghai Astronomical Observatory, United Kingdom Participation Group, Universidad Nacional Aut\'onoma de M\'exico, University of Arizona, University of Colorado Boulder, University of Oxford, University of Portsmouth, University of Utah, University of Virginia, University of Washington, University of Wisconsin, Vanderbilt University, and Yale University.

\end{document}